\documentclass[11pt]{article}
\usepackage{graphicx}
\usepackage{epsfig}
\usepackage{relsize}
\RequirePackage{xspace}

\newcommand{\AmS}{{\protect\the\textfont2
  A\kern-.1667em\lower.5ex\hbox{M}\kern-.125emS}}

\newcommand{\BABARPubYear}    {08}

\newcommand{\BABARProcNumber} {030}
\newcommand{\SLACPubNumber} {13527}

\def\babar{\mbox{\slshape B\kern-0.1em{\smaller A}\kern-0.1em
    B\kern-0.1em{\smaller A\kern-0.2em R}}}

\def\Dbar    {\kern 0.2em\overline{\kern -0.2em D}{}\xspace}
\def\Dz      {\ensuremath{D^0}\xspace}
\def\Dzb     {\ensuremath{\Dbar^0}\xspace}
\def\invfb   {\ensuremath{\mbox{\,fb}^{-1}}\xspace}

\begin{document}
\pagestyle{empty}

\begin{flushright}
SLAC-PUB-\SLACPubNumber \\
\babar-PROC-\BABARPubYear/\BABARProcNumber \\
January, 2009 \\
\end{flushright}

\par\vskip 4cm

% Title of the paper
\begin{center}
\Large \bf Recent charm mixing results from \babar, Belle, and CDF
\end{center}
\bigskip

\begin{center}
\large 
M. Charles\\
University of Iowa\\
SLAC M/S 35, 2575 Sand Hill Road, Menlo Park, CA 94025\\
(from the \babar\ Collaboration)
\end{center}
\bigskip \bigskip

% Abstract
\begin{center}
\large \bf Abstract
\end{center}
A summary of the results of several recent studies of charm mixing
is presented. A number of different methods were used, including
  the measurement of lifetime ratios for final states of different $CP$, 
  time dependence of wrong-sign hadronic decays,
  fits to time-dependent Dalitz plots, and 
  searches for wrong-sign semi-leptonic decays.
Taken together, they suggest mixing is of order 1\%.
The status of searches for indirect $CP$ violation is also reported.

\vfill
\begin{center}
Contributed to the Proceedings of the 
$5^{\mathrm{th}}$ International Workshop on e+e- collisions from Phi to Psi,
7--10 April 2008, Laboratori Nazionali di Frascati, Italy.
\end{center}

\vspace{1.0cm}
\begin{center}
{\em Stanford Linear Accelerator Center, Stanford University, 
Stanford, CA 94309} \\ \vspace{0.1cm}\hrule\vspace{0.1cm}
Work supported in part by Department of Energy contract DE-AC02-76SF00515.
\end{center}

\newpage

\section{INTRODUCTION}

There has been an abundance of experimental results on 
charm meson mixing recently, following the watershed
announcements by the \babar\ and Belle collaborations
that they had independently seen evidence for it
at the level of three standard
deviations~($\sigma$)~\cite{bib:ws:kpi:babar,bib:lifetimeratio:belle}.
This paper presents a summary of new measurements from
\babar~\cite{bib:babar_detector},
Belle~\cite{bib:belle_detector},
and CDF~\cite{bib:cdf_detector}, together
with a brief discussion of the mixing formalism and implications
for physics beyond the Standard Model (SM). In addition,
the CLEO collaboration has produced a
number of results relevant to charm mixing; these were discussed
in a separate talk at this conference~\cite{bib:jim_phipsi}.

\subsection{Mixing formalism}

The charmed mesons are produced as flavour eigenstates
\Dz and \Dzb. To obtain the time evolution of these states,
we express them as a linear combination of the eigenstates of the
Hamiltonian, $D_{1,2}$:
\begin{eqnarray}
  \label{eqn:defineD1}
  \left| D_1 \right\rangle &=& p \left| \Dz \right\rangle + q \left| \Dzb \right\rangle \\
  \label{eqn:defineD2}
  \left| D_2 \right\rangle &=& p \left| \Dz \right\rangle - q \left| \Dzb \right\rangle,
\end{eqnarray}
where $\left|q\right|^2 + \left|p\right|^2 = 1$ and
$CPT$ conservation is assumed.
The time evolution of the states $D_{1,2}$ is simply given by
the Schroedinger equation:
\begin{equation}
  \label{eqn:schroedinger}
  \left| D_{1,2} (t) \right\rangle = e^{-i(m_{1,2} - i\Gamma_{1,2}/2)t} \left| D_{1,2} (t=0) \right\rangle ,
\end{equation}
where $m_{1,2}$ and $\Gamma_{1,2}$ are the masses and widths of the
states $D_{1,2}$, respectively. Combining
equations~\ref{eqn:defineD1}--\ref{eqn:schroedinger},
we can invert the expressions above to obtain the
the time-dependence of the flavour eigenstates
$|\Dz\rangle$ and $|\Dzb\rangle$
in terms of the parameters $p$, $q$, $m_{1,2}$, and
$\Gamma_{1,2}$.
If $m_1 \neq m_2$ or $\Gamma_1 \neq \Gamma_2$ then mixing will occur:
a state which is initially composed only of \Dz will in general contain
a component of \Dzb at time $t > 0$, and vice versa. We quantify this effect
with the dimensionless mixing parameters $x$ and $y$, defined as:
\begin{eqnarray}
  &x = \frac{\Delta m}{\Gamma} = \frac{m_1 - m_2}{\Gamma} \\
  &y = \frac{\Delta \Gamma}{2\Gamma} = \frac{\Gamma_1 - \Gamma_2}{2\Gamma} ,
\end{eqnarray}
where $\Gamma = (\Gamma_1 + \Gamma_2)/2$.
We also define $R_M = (x^2 + y^2)/2$.

In the SM, mixing can occur through short-range box-diagram processes,
and through long-range rescattering processes via intermediate hadronic
states. The former are heavily suppressed by the GIM mechanism or by
$|V_{ub}V_{cb}|^2$, and are expected to be small in comparison
to the long-range contributions~\cite{Falk:2004wg}.
The latter are difficult to calculate precisely, with some recent
predictions for $x$ and $y$ of
order $10^{-2}$ to $10^{-3}$~\cite{Falk:2004wg}.
New physics (NP) beyond the SM could also contribute,
but unless the effect were
extremely large it would be obscured by current theoretical uncertainties
in the SM mixing rate. Experimental upper limits on the
mixing parameters have been used 
to constrain the parameter space of NP models~\cite{bib:hewett}.

The situation is quite different for $CP$ violation (CPV): 
SM contributions to direct CPV are expected to be small
($\mathcal{O}(10^{-3})$ or less, depending on the final state),
and indirect CPV should be negligible~\cite{bib:grossman_kagan_nir}.
Therefore, if CPV were observed in the charm system at the
present level of experimental sensitivity, it would be 
strong evidence for physics beyond the SM.

\section{EXPERIMENTAL RESULTS}
\label{sec:exptl}

\subsection{Common reconstruction and selection strategies}
\label{sec:exptl:cuts}

Charm mixing is a small effect. Therefore, each of the analyses
described below relies on having a very clean sample of correctly
reconstructed charm decays: significant levels of background
would wash out a mixing signal and ruin the sensitivity. The main
sources of background are combinatoric (e.g. tracks from light quark
jets) and mis-reconstructed charm decays. 
Secondary charm mesons produced in $B$ decays must either be
removed or have their production vertices measured accurately.

Background is suppressed by restrictions on the kinematic properties
of the \Dz candidate:
  the center-of-mass frame momentum,
  the measured proper lifetime and the associated uncertainty, and
  the invariant mass.
Particle identification (PID) requirements are also imposed on
its daughters. The \Dz candidate must also come from a reconstructed
$D^{*+} \to \Dz \pi^+$ or $D^{*-} \to \Dzb \pi^-$ decay and pass
corresponding kinematic selection criteria.

\subsection{Decays to \boldmath{$CP$} eigenstates}

If mixing is present, the decay time distributions of $D^0$ mesons
to final states with different $CP$ may vary. This effect is familiar
from the kaon system, where the mean lifetime of $K^0$ decaying to
$CP$-even states is very much shorter than for $CP$-odd states. The charm
system does not have such a dramatic difference, but with precision
studies of the large \babar\ and Belle data samples it is possible to
search for percent-level effects. The results from recent
searches by \babar~\cite{bib:lifetimeratio:babar}
and Belle~\cite{bib:lifetimeratio:belle}
are presented below.

Since the level of mixing is small, the expected decay time distribution
of \Dz (\Dzb) to a $CP$-even final state $h^+ h^-$ for $h=K,\pi$ can be
approximated as a single exponential with
mean lifetime $\tau^+_{hh}$ ($\tau^-_{hh}$).
We also approximate the distribution for the right-sign (RS) decay
$D^0 \to K^- \pi^+$ and its complex conjugate as a single exponential
with mean $\tau_{K\pi} = 1/\Gamma$. We can then define the mixing
parameter $y_{CP}$ and the $CP$ observables $A_{\Gamma}$ and
$\Delta Y$ as:
\begin{eqnarray}
      y_{CP} &=& \frac{\tau_{K\pi}}{(\tau^+_{hh} + \tau^-_{hh})/2} - 1 \\
  A_{\Gamma} &=& -\frac{\tau^+_{hh} - \tau^-_{hh} }{\tau^+_{hh} + \tau^-_{hh} } \\
    \Delta Y &=& -\frac{\tau_{K\pi}}{(\tau^+_{hh} + \tau^-_{hh})/2} A_{\Gamma}.
\end{eqnarray}
In the absence of CPV, $A_{\Gamma} = \Delta Y = 0$ and
$y_{CP} = y$. In the absence of mixing, all three parameters are zero.

Since the measured quantities are ratios of the
mean lifetimes of topologically identical and kinematically similar
final states, many systematic effects cancel. For example, in a
study of simulated events the \babar\ collaboration showed that misalignment
of the silicon vertex detector could introduce a bias of order 3~fs
(0.7\%) in the measured lifetimes---but that it was almost completely
correlated between the different final states and had only a small
effect (0.06\%) on $y_{CP}$ and $\Delta Y$.

Both Belle and \babar\ use strict selection criteria to suppress
background, as discussed in section~\ref{sec:exptl:cuts}. \babar\ also
places a requirement on the cosine of the helicity angle of the decay.
Background suppression is crucial to this analysis since the
composition of the backgrounds differs between
final states, and so the systematic uncertainties associated
with modelling the rate and time-dependence of background typically do
not cancel when taking ratios of lifetimes.

The results of the fit are
shown in Table~\ref{tab:lifetimeratio}. Both collaborations find
evidence for mixing (with statistical significance of $3.0\sigma$ and 
$3.2\sigma$ for \babar\ and Belle, respectively); neither
finds evidence for CPV. In addition, an older result from
\babar\ is shown~\cite{bib:lifetimeratio:oldbabar}:
the data sample used contained only $\Dz$ not tagged
with a $D^*$ decay and is statistically independent from 
the more recent analysis of tagged mesons.

\begin{table}[tb]
\caption{
  Fit results for $y_{CP}$ and for the $CP$-violating observable used
  ($\Delta Y$ for \babar\ and $A_{\Gamma}$ for Belle).
}
\label{tab:lifetimeratio}
\newcommand{\m}{\hphantom{$-$}}
\newcommand{\cc}[1]{\multicolumn{1}{c}{#1}}
\renewcommand{\tabcolsep}{1pc} % enlarge column spacing
\renewcommand{\arraystretch}{1.2} % enlarge line spacing
\begin{center}
\begin{tabular}{lcc}
\hline
\cc{Sample} & $y_{CP}$ & CPV \\
\hline
Belle tagged    (540~\invfb) & $(1.31 \pm 0.32 \pm 0.25)\%$ & $(+0.01 \pm 0.30 \pm 0.15)\%$ \\
\babar\ tagged  (384~\invfb) & $(1.24 \pm 0.39 \pm 0.13)\%$ & $(-0.26 \pm 0.36 \pm 0.08)\%$ \\
\babar\ untagged (91~\invfb) & $(0.8 \pm 0.4 ^{+0.5}_{-0.4})\%$ & $(-0.8 \pm 0.6 \pm 0.2)\%$ \\
\hline
\end{tabular}
\end{center}
\end{table}

\subsection{Wrong-sign hadronic decays}

The time-dependence of the decay rate to other final states is
also affected by mixing. The effect is clearest for wrong-sign (WS)
decays such as $\Dz \to K^+ \pi^-$, which in the absence of
mixing occur only via doubly-Cabibbo-suppressed (DCS) decays
with a small rate $R_D$
and have a pure exponential distribution of the form 
$\Gamma_{WS}(t) \propto R_D e^{-\Gamma t}$.
When mixing is allowed, a second mechanism opens up: an initially
pure \Dz state will evolve to include a component of \Dzb that
can undergo a Cabibbo-favoured (CF) decay to $K^+ \pi^-$.
Assuming that the mixing rate is small and neglecting $CP$ violation,
the time-dependence then becomes
\begin{equation}
  \label{eq:ws}
  \frac{\Gamma_{WS}(t)}{e^{-\Gamma t}} \propto
  R_D + 
  y^{\prime}\sqrt{R_D}(\Gamma t) +
  \frac{x^{\prime 2} + y^{\prime 2}}{4}(\Gamma t)^2
\end{equation}
where
  $x^{\prime} = x \cos \delta_{K\pi} + y \sin \delta_{K\pi}$,
  $y^{\prime} =-x \sin \delta_{K\pi} + y \cos \delta_{K\pi}$,
and $\delta_{K\pi}$ is the strong phase between the DCS and
CF amplitudes. By fitting this time-dependence, the
mixing parameters $x^{\prime 2}$ and $y^{\prime}$ can be extracted.
When allowing for CPV, the form of
Eq.~\ref{eq:ws} remains the same but has separate coefficients
$R_D^{\pm}$, $x^{\prime 2 \pm}$, and $y^{\prime \pm}$ for
\Dz decays $(+)$ and \Dzb decays $(-)$.

\babar, Belle, and CDF have carried out searches for mixing
in WS $\Dz \to K^+ \pi^-$
events in samples of 384, 400, and 1.5~\invfb of data,
respectively~\cite{bib:ws:kpi:babar,bib:ws:kpi:belle,bib:ws:kpi:cdf},
using the selection criteria described in section~\ref{sec:exptl:cuts}.
The mixing parameters obtained are shown in
Table~\ref{tab:ws:kpi}. \babar\ and CDF find mixing signals with
statistical significances of $3.9\sigma$ and $3.8\sigma$
respectively, indicating clear evidence for mixing. \babar\ and
Belle also searched for CPV, but found no evidence for
such an effect.

\begin{table*}[htb]
\caption{
  Results from fits to WS $\Dz \to K^+ \pi^-$ decays.
  The DCS rate $R_D$ and mixing parameters $y^{\prime}$ and
  $x^{\prime 2}$ are given, along with the statistical significance
  of the mixing signal in standard deviations. Where measured,
  the $CP$-violating asymmetry parameters $A_D$ and $A_M$ are given;
  these are the asymmetries in $R_D$ and $R_M$,
  respectively.
}
\label{tab:ws:kpi}
\newcommand{\m}{\hphantom{$-$}}
\newcommand{\cc}[1]{\multicolumn{1}{c}{#1}}
\begin{tabular}{lcccccc}
\hline
Experiment & $R_D (10^{-3})$ & $y^{\prime}$        &  $x^{\prime 2}$        & Signif. & $A_D$ (\%)             & $A_M$ \\
\hline
CDF (1.5~\invfb)       & $3.04 \pm 0.55$ & $8.5 \pm 7.6$       & $-0.12 \pm 0.35$       & $3.8$   &                        & \\
\babar\ (384~\invfb)     & $3.03 \pm 0.19$ & $9.7 \pm 5.4$       & $-0.22 \pm 0.37$       & $3.9$   & $-2.1 \pm 5.2 \pm 1.5$ & \\
Belle  (400~\invfb)    & $3.64 \pm 0.17$ & $0.6^{+4.0}_{-3.9}$ & $0.18^{+0.21}_{-0.23}$ & 2.0     & $2.3 \pm 4.7$          & $0.62 \pm 1.2$ \\
\hline
\end{tabular}
\end{table*}

The mixing parameters of other WS hadronic decays can be 
extracted from their time-dependence in a similar fashion.
With multi-body decays such as $\Dz \to K^+ \pi^- \pi^0$,
additional sensitivity can be gained by including the position
of each event within the Dalitz plot in the fit, since the
distributions of the DCS and CF decays differ. \babar\ has used
this technique in a search for mixing in
$\Dz \to K^+ \pi^- \pi^0$, obtaining a preliminary result of 
$R_M = (2.9 \pm 1.6) \times 10^{-4}$ from a sample of 384~\invfb.
%This is consistent with no mixing at the 0.8\% level.

\subsection{\boldmath{$D^0 \to K_S \pi^+ \pi^-$}}

As a special case of the analyses discussed above, final states
such as $K_S \pi^+ \pi^-$ contain several different classes of
contribution in the same Dalitz plot. These include 
  RS decays (e.g. $\Dz \to K^{*-} \pi^+$),
  WS decays (e.g. $\Dz \to K^{*+} \pi^-$),
  $CP$-even final states (e.g. $K_S \rho^0$), and
  $CP$-odd final states (e.g. $K_S f_0$).
All of these amplitudes are present in the same Dalitz plot
and interfere, allowing their relative phases to be determined
in the fit. As a result, the mixing parameters $x$ and $y$ can
be measured directly. The Belle collaboration has carried out
a search for mixing in this final state in 540~\invfb of
data~\cite{bib:kspipi:belle} and obtained:
\begin{eqnarray*}
  x &=& \left( 0.80 \pm 0.29 \, ^{+0.09}_{-0.07} \, ^{+0.10}_{-0.14} \right) \% \\
  y &=& \left( 0.33 \pm 0.24 \, ^{+0.08}_{-0.12} \, ^{+0.06}_{-0.08} \right) \% .
\end{eqnarray*}
They found no evidence for $CP$ violation.

\subsection{Semi-leptonic decays}

In contrast to the hadronic processes discussed above, for
which there are multiple contributions to the same final
state including a non-mixing component, wrong-sign semi-leptonic
decays $\Dz \to K^{(*)+} l^- \bar{\nu}_l$ can only occur
via mixing in the Standard Model---and so any observation
would be unambiguous evidence of mixing. However, the
experimental reconstruction and selection of these events is
made more difficult because the reconstructed \Dz is
incomplete---the neutrino and possibly additional $K^{*+}$
daughters are not found---and so 
the power of the kinematic selection criteria discussed in
section~\ref{sec:exptl:cuts} is reduced.

\babar\ and Belle both search for mixing in these final states,
but use quite different techniques.
\babar\ uses only $l=e$ and obtains a very pure sample with a
double-tag technique: in addition to the signal-side \Dz,
the recoiling charm decay in the opposite hemisphere of the
event is also reconstructed. This lowers the background
dramatically, but also
reduces the signal efficiency. They find no evidence for mixing
in a sample of 344~\invfb~\cite{bib:semilep:babar}
and set a 90\% confidence interval of 
$-13\times10^{-4} < R_M < 12\times10^{-4}$.
This is entirely consistent with the evidence for mixing discussed
above, which found $R_M \sim 10^{-4}$.

Belle use both the electron and muon decay modes and do not
require an opposite-side tag. Instead they apply additional
kinematic constraints to the $K^+ l^-$ system, and use
information on the missing energy and momentum in the event
to improve the reconstruction and allow more stringent
selection criteria. They find no evidence for mixing 
in a sample of 492~\invfb
and set a preliminary 90\% confidence limit of 
$R_M < 6.1 \times 10^{-4}$~\cite{bib:semilep:belle}.

\subsection{Combined results}

\begin{figure}[tb]
%\vspace{9pt}
\begin{center}
  \epsfig{file=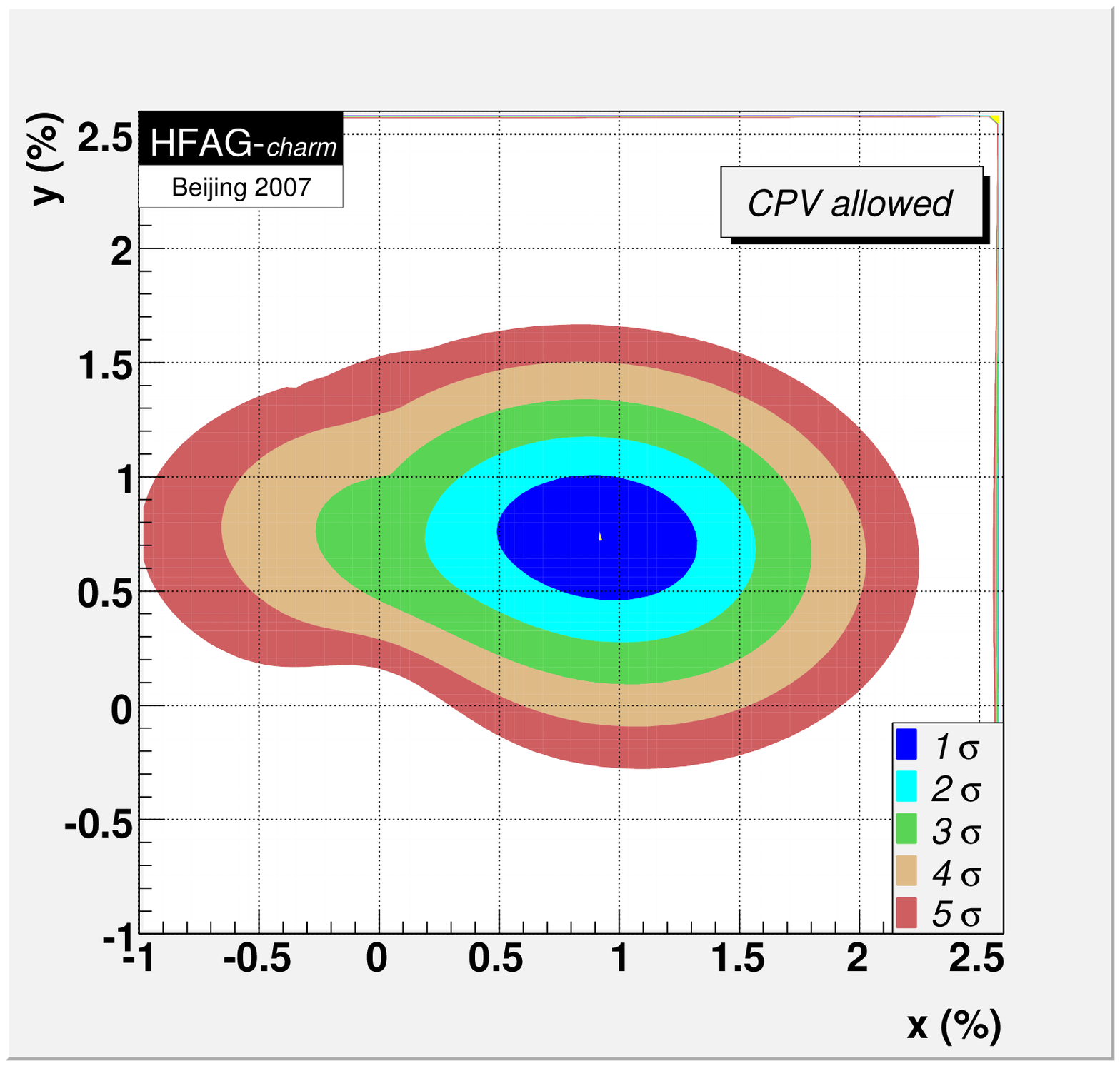, width=0.49\linewidth}
  \epsfig{file=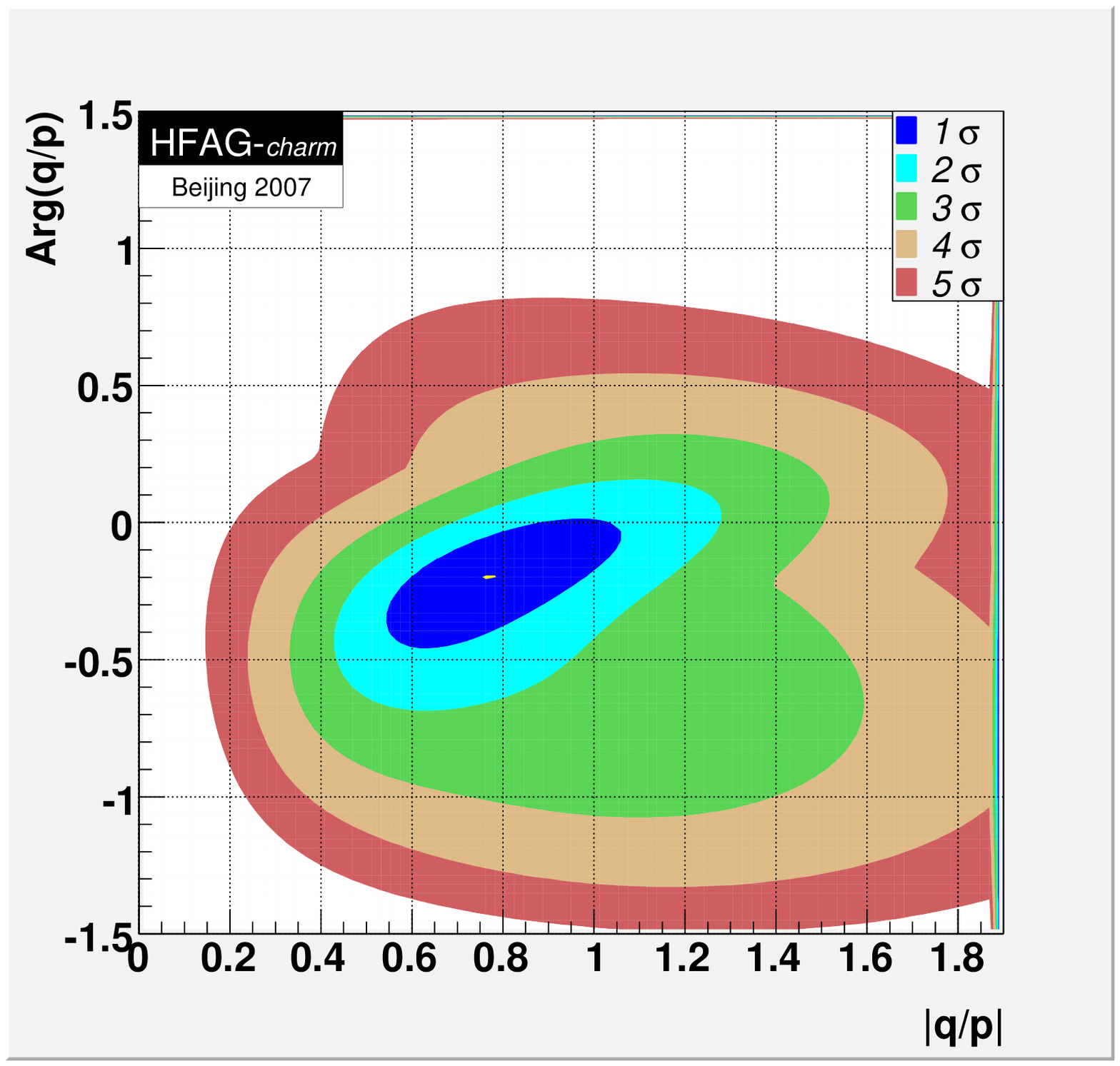, width=0.49\linewidth}
\end{center}
\caption{World-average results from the HFAG. The left plot shows
  limits on the mixing parameters $x$ and $y$ allowing for
  CPV, and the right plot shows the limits on the
  CPV parameters $\arg(q/p)$ and $|q/p|$.
}
\label{fig:hfag}
\end{figure}

By combining the above results with other relevant
measurements, the Heavy Flavor Averaging Group (HFAG)
has determined world-average values and confidence
intervals for the mixing and CPV parameters~\cite{bib:hfag}.
These are illustrated in Fig.~\ref{fig:hfag}.
Allowing CPV, they find:
\begin{eqnarray*}
  x &=& \left( 0.97 ^{+0.27} _{-0.29} \right) \% \\
  y &=& \left( 0.78 ^{+0.18} _{-0.19} \right) \% \\
  |q/p| &=& 0.86 ^{+0.18} _{-0.15} \\
  \arg(q/p) &=& \left( -0.17 ^{+0.14} _{-0.16} \right) \mathrm{rad}
  .
\end{eqnarray*}
The zero-mixing hypothesis ($x=y=0$) is excluded at the level
of $6.7\sigma$, but the data are still consistent
with zero indirect CPV ($|q/p|=1$, $\arg(q/p)=0$ is on
the $1\sigma$ contour).

\section{CONCLUSIONS}

Charm mixing is now established, with a combined
world significance of $6.7\sigma$.
However, no single measurement has yet exceeded
five standard deviations and there are large
uncertainties on the mixing parameters---there is
still more work to do, and we are still statistically
limited. The observed mixing rate is consistent
with the SM predictions, albeit within large
theory and experimental uncertainties, and 
appears to be at the upper end of the expected range.
There has been no sign of $CP$ violation, direct
or indirect, in the charm system yet. However,
limits on CPV are still well above the SM expectations
and there is plenty of room for new physics
to emerge in future measurements.

\end{document}